\documentclass[11pt]{article}
\pdfoutput=1
\usepackage{jheppub}
\usepackage{subfig}

\newcommand\beq{\begin{equation}}
\newcommand\eeq{\end{equation}}
\newcommand\be{\begin{equation}}
\newcommand\ee{\end{equation}}
\def\be{\begin{eqnarray}}
\def\ee{\end{eqnarray}}
\newcommand{\nn}{\nonumber}
\newcommand\para{\paragraph{}}

\newcommand{\eqn}[1]{(\ref{#1})}

\def\Dslash{\,\,{\raise.15ex\hbox{/}\mkern-12mu D}}
\def\Dbarslash{\,\,{\raise.15ex\hbox{/}\mkern-12mu {\bar D}}}
\def\delslash{\,\,{\raise.15ex\hbox{/}\mkern-9mu \partial}}
\def\delbarslash{\,\,{\raise.15ex\hbox{/}\mkern-9mu {\bar\partial}}}
\def\pslash{\,\,{\raise.15ex\hbox{/}\mkern-9mu p}}
\def\calDslash{\,\,{\raise.15ex\hbox{/}\mkern-12mu {\cal D}}}

\def\lae{\mathrel{\mathop{\smash{\lower .5 ex \hbox{$\stackrel<\sim$}}}}}
\def\lae{\mathrel{\mathop{\smash{\lower .5 ex \hbox{$\stackrel>\sim$}}}}}

\title{More  Abelian Dualities in 2+1 Dimensions}

\author[a]{Andreas Karch,}
\affiliation[a]{Department of Physics, University of Washington, Seattle, Wa, 98195-1560, USA}
\author[a]{Brandon Robinson}
\author[b]{and David Tong}
\affiliation[b]{Department of Applied Mathematics and Theoretical Physics, University of Cambridge, Cambridge, CB3 OWA, UK}

\preprint{\today}

\emailAdd{akarch@uw.edu, robinb22@uw.edu, d.tong@damtp.cam.ac.uk }

\abstract{We expand on the recent derivation of 3d  dualities using bosonization. We present in some detail a general class of Abelian duals.}

\begin{document}
\maketitle
%\tableofcontents

%%%%%%%%%%%%%%%%%

\section{Introduction}

In two spatial dimensions, flux attachment provides a simple way to transmute the quantum statistics of particles \cite{frank}. This is typically achieved by coupling particles to an emergent gauge field whose dynamics is governed by a Chern-Simons term \cite{arovas}.

\para
The original formulation of flux attachment involved non-relativistic matter. Extending this to relativistic matter is no small step. The presence of anti-particles means that we can no longer describe the physics in terms of the quantum mechanics of a finite number of degrees of freedom. Instead, the transmutation of quantum statistics becomes a much stronger statement about the dynamics of quantum field theory and is usually referred to as a {\it bosonization} duality.

\para
Proposals for  bosonization dualities were suggested long ago \cite{polyakov,cfw}. However, it took some time before precise statements of these dualities were made \cite{john,ofer}. These developments were informed, in part, by the progress in understanding non-Abelian bosonization dualities and their relation to higher spin theories   \cite{ Aharony:2011jz,Giombi:2011kc, Aharony:2012nh}.

\para
Here we restrict our attention to the Abelian case. One such bosonization duality can be schematically stated as
\be \mbox{free fermion\ \ \ $\longleftrightarrow$\ \ \ $U(1)_1$ coupled to a boson},\label{this}\ee
where $U(1)_k$ means that the boson is coupled to a $U(1)$ gauge field with Chern-Simons level $k$. Another form of the bosonization duality reads
\be \mbox{boson \ \ \ $\longleftrightarrow$\ \ \ $U(1)_{-1/2}$ coupled to a fermion}.\label{that}\ee
The precise, path integral description of the dualities, as reviewed in Section \ref{reviewsec}, require the inclusion of self-interactions for the scalars, together with couplings to background gauge fields.

\para
It was recently shown that one could use  either of the dualities \eqn{this} or \eqn{that}  as starting point to derive a web of further dualities \cite{Karch:2016sxi,Seiberg:2016gmd}. This is achieved by manipulating the path integrals using techniques previously developed in \cite{Kapustin:1999ha} in the context of supersymmetric theories where background sources  --- in particular, background gauge fields --- are promoted to dynamical fields. (See also \cite{Witten:2003ya}.) Application of this method gives a derivation of a number of other dualities including the familiar particle-vortex duality for bosons \cite{peskin,dh,fl} and the more recently discovered version for fermions \cite{Son:2015xqa,wang1,metlitski1,mross}. Further papers  exploring aspects of particle-vortex duality and bosonization include \cite{Murugan:2016zal,Hsin:2016blu,Radicevic:2016wqn,Kachru:2016rui,Filothodoros:2016txa,Kachru:2016aon}.

\para
The purpose of the current paper is  use these  same techniques to derive a general class of Abelian dualities \footnote{Abelian dualities in quiver theories have previously appeared in, e.g., \cite{PhysRevB.71.144508}.}, involving multiple gauge groups and multiple matter fields. The pattern of dualities that emerges looks very similar  to the dual theories one finds in supersymmetric gauge theories, where these dualties are known as ``mirror symmetry" \cite{Intriligator:1996ex}. Nonetheless, there are a number of subtleties in the construction of these general Abelian dualities, and it is instructive to spell them out explicitly.

\para
In Section \ref{reviewsec}, we review the underlying bosonization duality and some of its properties, including the emergent time-reversal invariance and the important role played by the scalar self-interactions. The next two sections then describe classes Abelian duals that can be derived from the basic bosonization duality, either without (Section \ref{moresec}) of with (Section \ref{evenmoresec}) invoking the hidden time-reversal invariance.

\section{A Review of Abelian Bosonization}\label{reviewsec}

Prior to constructing general Abelian dualities, we will first review the precise statement of the basic bosonization dualities, \eqn{this} and \eqn{that}.  The notation established in this section will serve as a template for the more general cases to follow.

\para
Roughly stated, bosonization relates the partition functions for a theory of bosons and a theory of fermions. We start with a complex scalar $\phi$ with quartic self-interactions, which flows to the Wilson-Fisher fixed point. Coupling $\phi$ to a background gauge field $A$, the action takes the form
\be S_{\rm scalar}[\phi;A] = \int d^3x\ |(\partial_\mu - i A_\mu)\phi|^2  - \alpha |\phi|^4\label{scalar}.\ee
%
%\be S_{\rm scalar}[\phi;A] = \int d^3x\ |(\partial_\mu - i A_\mu)\phi|^2  +\ldots \label{scalar}\ee
%
We flow to  the infra-red by sending $\alpha \rightarrow \infty$ and tuning the mass to zero in order to hit the critical point.
The partition function for the scalar field  is then defined to be
\be Z_{\rm scalar}[A] =  \lim_{\alpha\rightarrow\infty}\int {\cal D}\phi\ \exp\Big(iS_{\rm scalar}[\phi;A]\Big).\nn\ee
The action for a free, massless fermion is given by
\be S_{\rm fermion}[\psi;A] = \int d^3x\ i\bar{\psi}\gamma^\mu (\partial_\mu - iA_\mu)\psi \label{fermion},\ee
and correspondingly, the partition function
\be
Z_{\rm fermion}[A]  = \int {\cal D}\psi \ \exp\Big(iS_{\rm fermion}[A] \Big)\nn.\ee
This is more subtle to define because this partition function is not gauge invariant. To remedy this, the fermionic partition function should be accompanied by a Chern-Simons term with half-integer level\footnote{Our notation follows that of \cite{Karch:2016sxi}, where we explicitly denote the  half-integer Chern-Simons terms. This differs from \cite{Seiberg:2016gmd} where the fermionic determinant is defined with an implicit Chern-Simons term at level $1/2$.}, where the Chern-Simons action is defined by
\be  S_{CS}[A] = \frac{1}{4\pi}\int d^3x\ \epsilon^{\mu\nu\rho}A_\mu\partial_\nu A_\rho\label{csd}.\ee
The final ingredient that we will need is an interaction between two different gauge fields. This is achieved through a BF-coupling of the form
\be S_{BF}[A;B] = S_{BF}[B,A]= \frac{1}{2\pi} \int d^3x\ \epsilon^{\mu\nu\rho} A_\mu \partial_\nu B_\rho\label{bf}.\ee
Both $S_{CS}$ and $S_{BF}$ are normalised such that the $e^{i k S_{CS/BF}}$ is gauge invariant when $k\in {\bf Z}$. It will prove notationally useful to also define the coupling
\be S_{CS}[A;B] = \frac{1}{4\pi}\int d^3x\ \epsilon^{\mu\nu\rho}A_\mu\partial_\nu B_\rho\label{cs}\ee
While this differs from $S_{BF}$ only by a factor of 2, its utility comes from the fact that we will often be dealing with many gauge fields with couplings $\kappa^{ab} S_{CS}[A_a;A_b]$; this notation allows us to  avoid splitting  these into diagonal $S_{\rm CS}[A]$ terms and off-diagonal $S_{BF}$ terms.

\subsubsection*{The Simplest Bosonization Dualities}

We are now in a position to better describe the `seed duality', heuristically \eqn{this}, from which others can be derived. Written as an equality between partition functions, it becomes
\be  Z_{\rm fermion}[A] \, e^{-\frac{i}{2}S_{CS}[A]}  = \int {\cal D} a\  Z_{\rm scalar}[a]\,e^{i S_{CS}[a] + i S_{BF}[a;A]},\label{dual1}\ee
where $A$ is a background gauge field while $a$ is a dynamical gauge field.

\para
The equality of partition functions \eqn{dual1} should be interpreted as agreement between the two theories in the infra-red. There is no Maxwell term $\frac{1}{4e^2} (da)^2$ for the dynamical gauge field because the infra-red limit involves  $e^2\rightarrow \infty$ (while remaining smaller than any UV cut-off).

%\para
%It remains to specify which  $\ldots$ we should take in  the actions \eqn{scalar} and \eqn{fermion} so that the duality \eqn{dual1} holds. This was, apparently, first stated clearly in the context of the large $N$ non-Abelian dualities \cite{sjain}. There are two choices but, for now, we will pick the free fermion on the left-hand-side and the Wilson-Fisher scalar, with quartic interactions, on the right-hand side. The other choice will be discussed in Section \ref{rgsec}.

\para
To get a flavour of the kind of manipulations that we will perform in this paper, we will now explain how to derive the duality \eqn{that} starting from \eqn{dual1}. The basic idea is to promote the background gauge field $A$ to a new, dynamical gauge field which we denote as $\tilde{a}$. We introduce a new background field, $C$, which couples through $S_{BF}[\tilde{a},C]$. The equality of partition functions then becomes
\be \int {\cal D}\tilde{a}\ Z_{\rm fermion}[\tilde{a}] e^{-\frac{i}{2}S_{CS}[\tilde{a}] -iS_{BF}[\tilde{a},C]}  = \int {\cal D} a {\cal D}\tilde{a}\  Z_{\rm scalar}[a]\,e^{i S_{CS}[a] + i S_{BF}[a;\tilde{a}] -iS_{BF}[\tilde{a},C]}.\nn\ee
On the right-hand-side, we can integrate out $\tilde{a}$. This means that we replace it by its equation of motion, $da=dC$, which, in turn, freezes the other dynamical gauge field.  The upshot is that this duality becomes
\be\int {\cal D}\tilde{a}\ Z_{\rm fermion}[\tilde{a}] e^{-\frac{i}{2}S_{CS}[\tilde{a}] -iS_{BF}[\tilde{a},C]}  =   Z_{\rm scalar}[C]\,e^{i S_{CS}[C]}.\label{dual2}\ee
This is the more precise statement of the duality \eqn{that}; it relates the Wilson-Fisher boson on the right-hand-side to a fermionic Chern-Simons-matter theory on the left-hand-side.

\subsubsection*{Time Reversal}

There is surprising feature of the basic bosonization dualities, \eqn{dual1} and \eqn{dual2}: one side exhibits manifest time reversal symmetry while the other does not. Time reversal invariance must then arise  as a hidden quantum symmetry on the sides of the dualities  with dynamical Chern-Simons terms.

\para
We can make use of this.  We can act with time reversal on both sides of \eqn{dual1} and \eqn{dual2}, flipping the sign of any Chern-Simons and BF couplings This results in two new dulaities. Starting from \eqn{dual1}, we get
\be  Z_{\rm fermion}[A] \, e^{+\frac{i}{2}S_{CS}[A]}  = \int {\cal D} a\  Z_{\rm scalar}[a]\,e^{-i S_{CS}[a] - i S_{BF}[a;A]}.\label{dual3}\ee
Whereas applying time reversal to \eqn{dual2}, we get
\be\int {\cal D}\tilde{a}\ Z_{\rm fermion}[\tilde{a}] e^{+\frac{i}{2}S_{CS}[\tilde{a}] +iS_{BF}[\tilde{a},C]}  =   Z_{\rm scalar}[C]\,e^{-i S_{CS}[C]}.\label{dual4}\ee
It was shown in \cite{Karch:2016sxi,Seiberg:2016gmd}  that this hidden time reversal invariance, and the resulting duals \eqn{dual3} and \eqn{dual4}, is what allows the derivation of particle-vortex duality from bosonization. This feature of the duality web will be explored further in Section \ref{evenmoresec}.

\subsection{Dualities and RG flow}\label{rgsec}

The partition functions in \eqn{dual1} depend on the background gauge field $A$. This background field couples to the currents  which are the easiest operators to identify on both sides of the duality.  In general, we would like include sources for all operators but the complete map between the two theories is difficult to determine. In this section, we will conjecture a map between the lowest dimension operators on the two sides. Although this conjecture is somewhat naive, it will help shed light on dual RG flows.

\para
We start with the bosonic side of the duality where there is an alternative way to write the action for a Wilson-Fisher scalar. We introduce
an real, auxiliary scalar $\sigma$ and consider the action for the pair of scalars,
\be S_{\rm scalar}[\phi;A;\sigma] = \int d^3x\ |(\partial_\mu - i A_\mu)\phi|^2  - \sigma |\phi|^2  + \frac{1}{2\alpha}\sigma^2.\nn\ee
If $\sigma$ is dynamical, we can integrate it out to take us back to our original action \eqn{scalar}. In the infra-red,  $\alpha\rightarrow \infty$ limit, we lose the $\sigma^2$ term, leaving $\sigma$ as a Lagrange multiplier.  This motivates us to define the more general partition function
\be Z_{\rm scalar}[A;\sigma] =  \lim_{\alpha\rightarrow \infty} \int {\cal D}\phi\ \exp\Big(iS_{\rm scalar}[\phi;A;\sigma]\Big).\nn\ee
On the level of the partition function, integrating out the dynamical $\sigma$ recovers the theory for the Wilson-Fisher scalar: $Z_{WF}[A] = \int {\cal D}\sigma\ Z_{\rm scalar}[A;\sigma]$. Alternatively, we could think of $\sigma$ as a background field, then it plays the role of source for the operator $|\phi|^2$ in $Z_{\rm scalar}[A;\sigma]$.

\para
With this in mind, let us now return to the question of the map between operators on the two sides of the duality \eqn{dual1}. We know that the fermion $\psi$ corresponds to a monopole operator in the Chern-Simons theory. The next simplest operator is $\bar{\psi}\psi$. A naive guess for the map is
\be \bar{\psi}\psi \ \ \ \longleftrightarrow\ \ \ -\sigma.\label{opmap}\ee
A similar map is known to hold in the large $N$ limit of non-Abelian bosonization dualities  where further mixing is $1/N$ suppressed  (see, for example, \cite{sjain,guy3}). However, there is no such protection against further, and likely, operator mixing in Abelian dualities. Nonetheless, below we will assume that map holds and see that it leads to a nice prediction for the RG flow of the dualities.

\para
Before we proceed, there is one obvious objection to the map \eqn{opmap}. The operator $\bar{\psi}\psi$  is odd under parity and time-reversal while one might be tempted to state that the operator $\sigma$ is even. However, the bosonic theory comes with a Chern-Simons term and, as described above, does not enjoy a manifest parity or time reversal symmetry. For this reason, we cannot easily assign a quantum number to $\sigma$. We note, however, that if $\sigma$  is indeed odd under the quantum symmetry then that ensures that any corrections to the map \eqn{opmap} starts at order ${\cal O}(\sigma^3)$.

\para
Let us look at the consequence of this operator map. We define the partition function for a massive fermion
\be Z_{\rm fermion}[A;m] = \int {\cal D}\psi\ \exp\Big(i\int d^3x\ i\bar{\psi}\gamma^\mu(\partial_\mu- iA_\mu)\psi - m\bar{\psi}{\psi} \Big).\nn\ee
The more refined version of the duality \eqn{dual1} now equates this to
\be  Z_{\rm fermion}[A;m] \, e^{-\frac{i}{2}S_{CS}[A]}  = \int {\cal D} a {\cal D}\sigma\  Z_{\rm scalar}[a;\sigma]\,e^{i (S_{CS}[a] +  S_{BF}[a;A] +m\sigma)}.\label{dual5}\ee
Note that the minus sign in \eqn{opmap} ensures that the Hall conductivities agree on both sides. This is given by $\sigma_{xy} = k/2\pi$, where $k$ is the infra-red Chern-Simons level for the background field $A$. When $m<0$, we can integrate out the fermion on the left-hand-side to leave behind a Hall conductivity $\sigma_{xy} = -1/2\pi$. On the bosonic side, the scalar is massive and can be integrated out;
subsequently integrating out $a$ reproduces this Hall conductivity. In contrast, when $m>0$, integrating out the fermion leaves us with vanishing Hall conductivity, $\sigma_{xy}=0$. The bosonic side mimics this in a rather nice way \cite{Aharony:2012nh,ofer}: integrating out $\sigma$ causes the scalar fields to condense, breaking the gauge group and leaving us with vanishing Hall conductivity.

\para
The theme of this paper is to derive new dualities from old by promoting background fields to dynamical fields. This method is not restricted only to background gauge fields. Indeed, the technique was originally developed in the context of supersymmetric theories for entire multiplets of background fields \cite{Kapustin:1999ha}. Here we can see what happens when we import this logic and promote the mass $m$ to a dynamical field which we rename $\mu$. The two sides of the duality now become
\be
\int {\cal D}\mu \ Z_{\rm fermion}[A;\mu] \, e^{-\frac{i}{2}S_{CS}[A] - i\mu \tilde{m}} &=& \int {\cal D} a {\cal D}\sigma {\cal D}\mu \  Z_{\rm scalar}[a;\sigma]\,e^{i (S_{CS}[a] +  S_{BF}[a;A] + \mu( \sigma -\tilde{m}))}
\nn\\ &=& \int {\cal D} a \ Z_{\rm scalar}[a;\tilde{m}]\,e^{i (S_{CS}[a] +  S_{BF}[a;A] )},
\nn\ee
where we have introduced  a new background source $\tilde{m}$. The right-hand side is now a regular scalar, without quartic interactions, coupled to a $U(1)_1$ Chern-Simons gauge field. Meanwhile, the left-hand-side is the infra-red limit of a Yukawa-type theory, with $\mu$ a real auxiliary scalar interacting with the fermion.

\para
There has been quite a lot of work in understanding the fixed point of these fermion-scalar theories with Yukawa-type interactions. For large number of flavors, $N_f$, the theory is known to describe a new conformal fixed point, usually referred to as the Gross-Neveu fixed point since it can be thought of as strongly deforming the action by the addition of quartic fermion terms. (See, for example  \cite{Rosenstein:1988pt} for a discussion of this theory.) However, it remains an outstanding problem to establish the existence of the Gross-Neveu fixed point for a small number of flavors.  While recent work utilizing conformal bootstrap in three dimensions has given evidence for the fixed point existing down to $N_f=1$ \cite{Pufu:2015qra}, this is still an open area of investigation.

\para
We see that the operator map \eqn{opmap}, if true, implies a second class of bosonization dualites in which the Gross-Neveu fermion is related to the free scalar. This behaviour is known to occur in large $N$ non-Abelian dualities \cite{sjain}; it is less clear if it extends to these Abelian dualities with $N_f=1$.

\para
If the Gross-Neveu fixed point does exist for $N_f=1$, it is expected that there is a relevant operator which initiates an  RG flow to the free, massless fermion. Relatedly, the free scalar can flow to the Wilson-Fisher fixed point. This suggests the RG flow between dualities is given by
%
%\begin{align}\nonumber
%\qquad\quad\text{UV}\qquad\qquad&\qquad\qquad\qquad \quad \text{IR}
%\\ \nonumber
%\text{Free Scalar} \qquad&\ridiculousrightarrow\qquad \text{WF Scalar}
%\\ \nonumber
%\updownarrow    \qquad\qquad&\qquad\qquad\qquad \quad\updownarrow
% \\ \nonumber
 % \text{GN Fermion}\qquad&\ridiculousrightarrow\qquad\text{Free Scalar}
%\end{align}
%
\begin{align} UV: \qquad\qquad \text{Free Scalar}\qquad\qquad & \longleftrightarrow\qquad\qquad \text{GN Fermion} \nn \\
\Big\downarrow\qquad\qquad \qquad & \qquad\qquad\qquad \qquad\quad\Big\downarrow \nn\\
IR\ : \qquad\qquad \text{WF Scalar}\qquad\qquad & \longleftrightarrow\qquad\qquad \text{Free Fermion} \nn \\
\nonumber \end{align}
Again, this mimics the situation in large $N$ non-Abelian theories.

\section{More Abelian Duals}\label{moresec}

In this section, we will derive a class of general Abelian dualities. Specifically, we start from the initial bosonization duality \eqn{dual1} without invoking its time-reversed partner  \eqn{dual3}.  Deriving dualities that require both \eqn{dual1} and \eqn{dual3} will be addressed in Section \ref{evenmoresec}.

\para
We begin by taking $N$ copies of the seed duality \eqn{dual1}. One side of the duality simply consists of $N$ free fermions with a $U(N)$ global symmetry, which is broken to $U(1)^N$ by the coupling to the background gauge fields $A_i$ where $i=1,\ldots,N$. We choose to gauge $U(1)^r\subset U(1)^N$ of these symmetries. The linear combination of fields is specified by the choice of fermion charges $R_i^a$ where $a=1,\ldots,r$ and $i=1,\ldots,N$. We take $R_i^a$ to have maximal rank. It is useful to also define the orthogonal set of charges $S_i^p$ with $p=1,\ldots,N-r$ which characterise the remaining global symmetries. These charges obey the constraint
\beq
\label{rs}
\sum_{i=1}^N R_i^a \,S_i^p =0,\ \ \ \ \ \forall\ a=1,\ldots,r\ {\rm and}\ p=1,\ldots,N-r.
\eeq
The gauge fields $A_i$ are decomposed according to the above splitting as
\beq
A_i = R_i^a \tilde{a}_a + S_i^p C_p,\label{adecompose}\eeq
where $\tilde{a}_a$ are the newly dynamical gauge fields and $C_p$ are the remaining background gauge fields. Relatedly, we decompose the $U(1)^N$ fermionic currents $(J^\mu)^i=\bar{\psi}^i\gamma^\mu\psi^i$ into the gauge currents $J^a= R_i^aJ^i$ and global currents $J^p=S^p_iJ^p$. Note that we are employing the useful compact, but slightly opaque, notation that different objects are distinguished only by the index they carry. The fermionic action then becomes
\be S_A &=&  i\bar{\psi}^i\!\delslash \psi^i + J^a\tilde{a}_a + J^pC_p  -\frac{1}{2} \kappa^{ab}S_{CS}[ \tilde{a}_a; \tilde{a}_b] - \frac{1}{2} \kappa^{pq} S_{CS}[C_p; C_q] -\kappa^{ab} S_{BF}[\tilde{C}_a; \tilde{a}_b], \ \ \ \ \
\label{ffinal}\ee
where we have introduced new background fields $\tilde{C}_a$ to couple to the topological currents, and defined the mixed Chern-Simons couplings
\beq
\kappa^{ab} =  \sum_i R^a_i R^b_i, \quad \quad \kappa^{pq} =  \sum_i S^p_i S^q_i.
\label{kappaab}\eeq
Spelled out in words, the resulting fermionic theory is:
\para
{\bf Theory A:}  $U(1)^r$ gauge theory with $N$ fermions of charge $R_i^a$ and Chern-Simons levels given by
$-\frac{1}{2} \kappa^{ab}=-\frac{1}{2} \sum_i R^a_i R^b_i$.

\vskip10pt

%\para
%Let us define will see that this $\kappa^{ab}$ gives us right normalization of the new BF term.
%With this the action of ``model A"  is given by
%\beq
%\nonumber S_A = \sum_i S_{fermion}[\psi_i;A_i] - \frac{1}{2} \sum_i S_{CS}[A_i] - \kappa^{ab} S_{BF}[ C_a; A_b]
%\eeq
%where $r$ linear combinations $A_a$ of the $N$ $A_i$ are gauged.
%
%
%\para
% For the manipulations below we also need to break up the fermion action into
%a kinetic part and a coupling of the gauge field to the current:
%\beq S_{fermion}[\psi_i;A_i] = S_{kin,f}[\psi_i] + S_{int}[J^iA_i] \eeq
%with
%\beq
%S_{int} [JA] = \int d^3x\, \sum_i  J^{\mu} A_{\mu}. \eeq
%Further let us define
%\beq
%A_i = R_i^a A_a + S_i^p A_p, \quad \quad J^a = R_i^a J^i, \quad \quad J^p = S_i^p J^i
%\eeq
%that is we broke up the original background fields $A_i$ into the newly dynamical fields $A_a$ and the orthogonal $A_p$ which remain background fields. Contracted indices are summed over as usual.
%With this we can nicely rewrite the action of model A as
%\begin{eqnarray} \nonumber S_A &=&
%\sum_i S_{kin,f}[\psi_i] +S_{int}[J^a A_a] + S_{int}[J^pA_p] \\ && -\frac{1}{2} \kappa^{ab}S_{CS}[ A_a; A_b] - \frac{1}{2} \kappa^{pq} S_{CS}[A_p; A_q] -\kappa^{ab} S_{BF}[C_a; A_b] .
%\end{eqnarray}
%or spelled out in words
%
%\vskip10pt
%

\para
Let us see where this same process takes us  on the bosonic side of the duality. We start with $N$ copies of the Wilson-Fisher scalars coupled to $U(1)_1^N$ Chern-Simons terms. Note that claimed duality already exhibits a surprising feature: even in the absence of background fields, this theory does not enjoy a manifest  $U(N)$ global symmetry but merely $U(1)^N$ topological symmetries.

\para
We perform the same gauging of $U(1)^r\subset U(1)^N$ as on the fermionic side. We similarly decompose the background fields $A_i$ as \eqn{adecompose} and promote $\tilde{a}_a$ to dynamical fields. It is convenient to perform a similar decomposition on the original dynamical gauge fields $a_i$ and scalar currents $(j^\mu)^i = i\phi^{i\,\dagger}{\partial}^\mu\phi^i - i\partial^\mu\phi^{i\,\dagger}\phi^i$ which we write as
\be
\label{bgaugedef}
a_i = R_i^a a_a + S_i^p a_p, \quad \quad j^a = R_i^a j^i, \quad \quad j^p = S_i^p j^i,
\label{olddecomp}\ee
Finally, as in the fermionic case, we  explicitly decompose the scalar action $S_{\rm scalar}[\phi;a]$ into the terms with dependence on the gauge field. We write
\beq
S_{\rm scalar}[\phi_i;a_i] = S_{0}[\phi_i] + S_{\rm int}[j^ia_i].
\eeq
where $S_0$ includes both kinetic terms and quartic couplings, while $S_{\rm int}$ includes the terms linear and quadratic in $a_i$. After this deluge of new notation, our bosonic theory becomes
\begin{eqnarray}
S_B &=& \sum_i S_0[\phi_i] +S_{\rm int}[j^a a_a]+S_{\rm int}[j^p a_p]
 + \kappa^{ab} S_{CS}[a_a; a_b]  + \kappa^{pq} S_{CS}[a_p; a_q] \nn \\&&\ \ \ \ \ \ \ \ \ \ \ \ \
 +\,  \kappa^{ab} S_{BF}[a_a -\tilde{C}_a;\tilde{a}_b] + \kappa^{pq} S_{BF}[ a_p; C_q],
\label{fullbaction} \ee
where, $a_a$, $a_p$ and $\tilde{a}_a$ are $N+r$ dynamical gauge fields, while  $C_p$ and $\tilde{C}_a$ are $N$ background fields.

\para
The full set of $N+r$ dynamical gauge fields in this theory is overkill. The newly promoted fields $\tilde{a}_a$ appear linearly in the action and can be integrated out. (There is a caveat to this statement involving flux quantisation which will be addressed below.) This, in turn, places constraints on the fields $a_a$. The upshot is that we are left with a theory possessing a $U(1)^{N-r}$ gauge symmetry. This is a story familiar in the supersymmetric theories \cite{Kapustin:1999ha} and plays out in much the same way here.
In detail, the $a_a$ and $\tilde{a}_a$ equations of motion read respectively
\be  *j^a + \frac{\kappa^{ab}}{2 \pi} (f_b +  \tilde{f}_b) =0\ \ \ {\rm and}\ \ \
 \kappa^{ab}(f_b  - d\tilde{C}_b) =0\label{newtwo}
\ee
Our assumption that $R_i^a$ had maximum rank means that $\kappa^{ab}$ is invertible and so these are solved by
\beq
a_a =  \tilde{C}_a, \quad \quad d\tilde{a}_a = -2 \pi \kappa_{ab} \, *j^b -  d\tilde{C}_a.
\nn\eeq
Plugging these back into the action, the $S_{BF}[a_a; \tilde{a}_b]$ and $S_{BF}[\tilde{C}_a; \tilde{a}_b]$ terms cancel and we are left with the bosonic theory
\be
S_B &=& \sum_i S_0[\phi_i] +S_{\rm int}[j^a C_a]+S_{\rm int}[j^p a_p]
 + \kappa^{pq} S_{CS}[a_p; a_q] \nn\\ &&\ \ \ \ \ \ \ \ \ \  \ \ \ \ \ \ \ \ \ \ \ \ \ \ \ \ \ \  +\, \kappa^{ab} S_{CS}[\tilde{C}_a; \tilde{C}_b]
 + \kappa^{pq} S_{BF}[ a_p; C_q]
 \label{bfinalanswer}
\ee
In words, this is:

\para

{\bf Theory B:}  $U(1)^{N-r}$ gauge theory with $N$ Wilson-Fisher scalars of charge $S_i^p$ and Chern-Simons levels
$\kappa^{pq}=\sum_i S^p_i S^q_i$.

\vskip10pt

\subsection*{A Comment on Flux Quantization}

The act of integrating out gauge fields using the equations of motion \eqn{newtwo} appears innocuous enough. And, indeed, for many local, dynamical questions the answers one derives from the bosonic theory \eqn{bfinalanswer} should coincide with those of the fermionic theory \eqn{ffinal}. However, there is one subtlety that the action \eqn{bfinalanswer} misses: flux quantisation.

\para
If handed \eqn{bfinalanswer}, one might think that in order to determine the allowed fluxes in Theory B one simply should impose Dirac quantization conditions for the $N-r$ remaining gauge fields. However, this would in general be incorrect, as the equations of motion \eqref{newtwo} need not be consistent with generic fluxes that are allowed from \eqref{bfinalanswer}.

\para
A similar issue has arisen recently in the discussion of particle-vortex duality where, in order to avoid the ${\bf Z}_2$ gauge anomaly (usually called the parity anomaly) one is obliged to restrict fluxes to twice their usual value \cite{Son:2015xqa,wang1,metlitski1}. Rather than impose this restriction by fiat, it was shown recently that it naturally arises by the addition of a second, auxiliary gauge field with suitable couplings \cite{Seiberg:2016gmd}. This is reminiscent of the way that dynamical, emergent gauge fields are needed to correctly describe fractional Hall conductivities.

\para
This same lesson carries over to the present discussion. We should not impose canonical flux quantisation on \eqn{bfinalanswer}. Instead,
 when faced with such global questions, we should retreat to the $U(1)^{N+r}$ gauge theory \eqn{fullbaction} where the canonical  flux quantization holds,
\beq
\label{quantization}
\int_{S^2} f_i, \, \int_{S^2} \tilde{f}_a \, \in  \, 2\pi{\bf Z}
\eeq
In contrast, for questions of local dynamics, the $U(1)^{N-r}$ action \eqn{bfinalanswer} should suffice.

\subsection{An example: QED$_3$ with a Chern-Simons Term}\label{qedsec}

\begin{figure}[t]
\begin{center}
\includegraphics[width=3.5in]{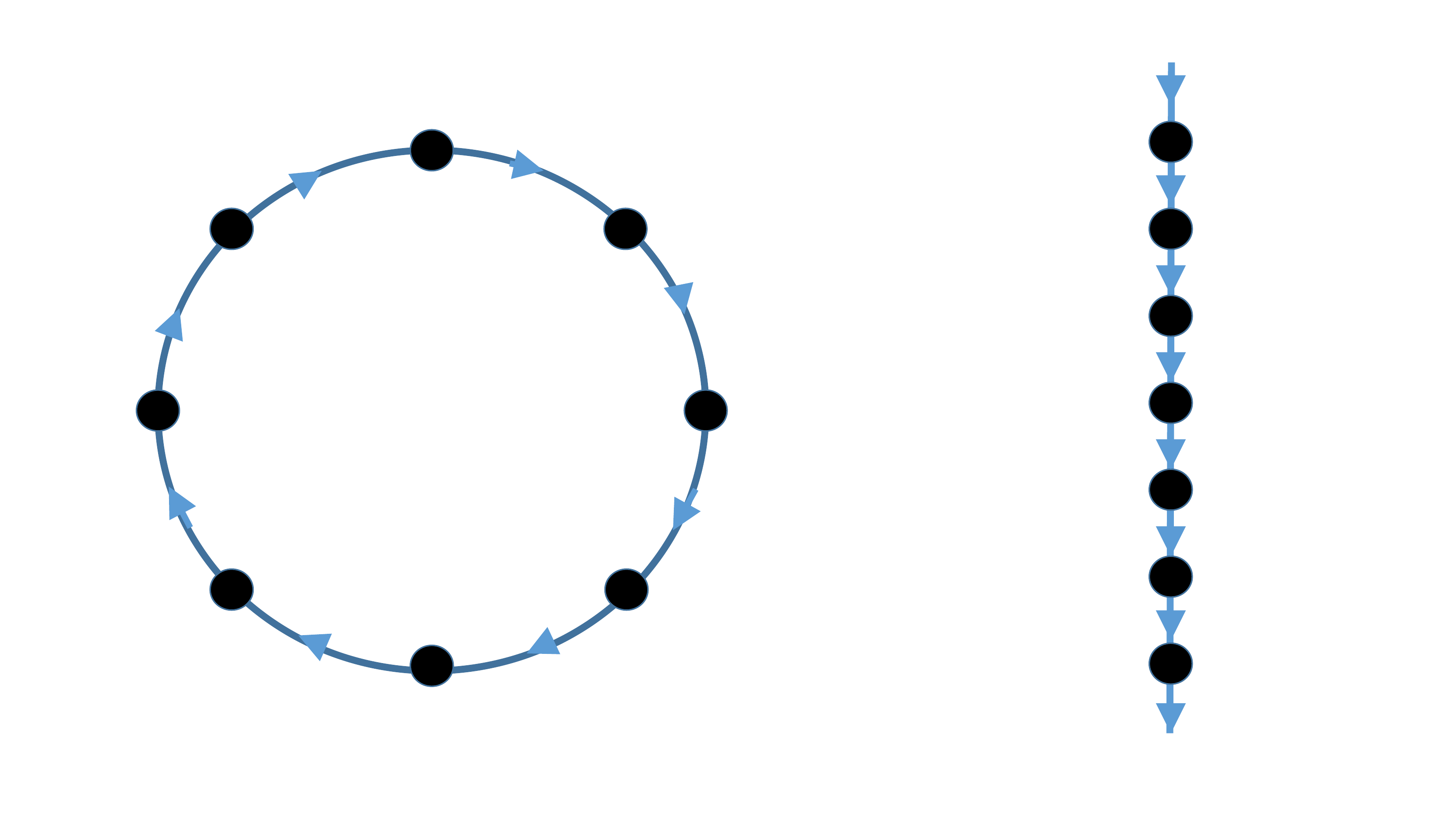}
\end{center}
\vskip -0.7cm \caption{The two different quiver representations of the $U(1)^7$ gauge theory associated with $N=8$. Nodes are $U(1)$ gauge groups, arrows represent matter with charge +1 under the node the arrow points to and -1 under the gauge group it originates from. In the circular quiver on the left the overall $U(1)$ (under which no matter is charged) still needs to be modded out. The circular quiver makes manifest a $Z_8$ permutation symmetry. The linear quiver on the right hand side only directly displays the 7 independent gauge groups at the price of not making manifest the discrete $Z_8$ symmetry. We mostly work with the linear quiver representation.}
\label{quiversfig}
\end{figure}

As a specific example, consider $r=1$ with $R^1_i=1$ for $i=1$ to $N$. The fermionic theory is QED with $N$ charged fermions and a Chern-Simons term:

\para {\bf Theory A:} $U(1)$ with $N_f=N$ fermions and Chern-Simons term at level $k=-N/2$.

\para

The dual, bosonic theory is a $U(1)^{N-1}$ quiver theory.  The name {\it quiver} comes from the mathematical literature and derives from the graphical representations of these theories. The standard picture for this quiver theory is $N$ $U(1)$ gauge groups arranged around a circle with bi-fundamental scalar matter with charge $(+1,-1)$ under neighbouring groups. The matter is neutral under the diagonal $U(1)$, ensuring that the full gauge symmetry is actually $U(1)^{N-1}$. This particular pair is a non-supersymmetric version of the very first supersymmetric mirror symmetry proposed in \cite{Intriligator:1996ex}. The graphical representation of this circular quiver for $N=8$ is shown in the left panel of figure \ref{quiversfig}.

\para
Alternatively, we can describe the theory directly without any redundancy. In this presentation, the dual bosonic theory is:

\para
{\bf Theory B:} $U(1)^{N-1}$ coupled to $N$ Wilson-Fisher scalars with charges $S^i_p$ ($i=1,\ldots,N$ and $p=1,\ldots,N-1$)
\beq \label{quivers}
S^1_i=\begin{pmatrix} +1 \\ -1 \\ 0 \\ \vdots \\ 0\\ 0 \end{pmatrix}, \quad
S^2_i=\begin{pmatrix} 0 \\ +1 \\ -1 \\ \vdots \\0 \\ 0 \end{pmatrix}, \quad
\ldots, \quad
S_i^{N-1}=\begin{pmatrix} 0 \\ 0 \\ 0 \\ \vdots \\ +1 \\-1 \end{pmatrix}.
\eeq
The Chern-Simons level is  $+2$ for each gauge group and $-1$ between neighbouring gauge groups such that the level matrix is given by
\be k^{pq} = \left(\begin{array}{ccccc} 2 & -1 & & & \\ -1 & 2 & -1 & \\ & - 1&  \ddots &  &  \\ & &  &  2 & -1  \\ & &  & -1 & 2
 \end{array}\right).\label{cartan}\ee
Note that $k^{pq}$ coincides with the Cartan matrix of $SU(N)$.

\para
Our charge matrices can be graphically represented as a linear quiver with open ends. In this case, the first and last matter multiplet are only charged under one of the nodes. For example, consider the $N=8$ linear quiver displayed in the right panel of figure \ref{quiversfig}. The disadvantage of the linear quiver is that it does not make manifest the explicit $Z_N$ symmetry of the original represntation that cyclically permutes the nodes.

\para
We note that a different dual for Theory A was previously conjectured by Aharony \cite{ofer}. This follows from the proposal
\be U(k)_{-N+N_f/2} \ \mbox{coupled to $N_f$ fermions}\ \ \ \longleftrightarrow\ \ \ \ SU(N)_k\ \mbox{coupled to $N_f$ scalars}\nn\ee
and restricting to $N=N_f$ and $k=1$. In general, these dualities with $N_f>1$ are not as well tested as those with a single flavour and some concerns were raised in \cite{Hsin:2016blu}. However, the particular choice $N_f=N$ was recently studied in \cite{Radicevic:2016wqn} where compelling evidence was found for the duality, at least in a non-relativistic limit. This suggests that the Theory B described above may itself be equivalent to a non-Abelian bosonic theory

\para
{\bf Theory B$'$:} $SU(N)_1$ coupled to $N_f$ fundamental scalars with quartic coupling
\para

\subsubsection*{Another Example: Scalar QED$_3$ with a Chern-Simons Term}

It is a simple matter to extend the above discussion to scalar QED. This time, we take $r=N-1$ with the scalar charges given by $S_i^1=1$ for all $i$. The bosonic theory then becomes

\para
{\bf Theory B:} $U(1)$ coupled to $N$ scalars with quartic couplings and Chern-Simons level $k=N$.

\para
Meanwhile, the fermionic theory is a $U(1)^{N-1}$ quiver whose matter charges $R^a_i$ are given by the vectors \eqn{quivers}. The dual is then

\para
{\bf Theory A:} $U(1)^{N-1}$ coupled to $N$ fermions with charges \eqn{quivers} and Chern-Simons levels $-\frac{1}{2}\kappa^{pq}$ with $\kappa^{pq}$ given in \eqn{cartan}.

\para
Once again, there is a different, non-Abelian dual proposed in \cite{ofer}. This time it follows from the duality:
\be U(N)_{k} \ \mbox{coupled to $N_f$ scalars}\ \ \ \longleftrightarrow\ \ \ \ SU(k)_{-N+N_f/2}\ \mbox{coupled to $N_f$ fermions}\nn\ee
Restricting to $N=1$ and $k=N_f$, this suggests that the Abelian fermionic quiver is equivalent to

\para {\bf Theory A$'$:} $SU(N)_{N/2-1}$ coupled to $N$ fermions

\para
Clearly, it would be interesting to understand the extent to which these proposed Abelian and non-Abelian duals are equivalent.

\subsection{Matching the Operator Spectrum}\label{matchingsec}

In order to  verify our conjectured duality, we would like to see matching between the operators on the two sides. Both  sides of the duality are strongly coupled which means that it is difficult to get a handle on the dimensions of operators.  Here, instead, we limit our ambition: we look at the quantum numbers of simple operators on both sides of the duality and check that they match. We will discuss the specific QED dualities described in Section  \ref{qedsec}.

\para
We start with Theory A described in the previous section:  QED with $N$ flavors and Chern-Simons level $-N/2$. The theory has an $U(1)\times SU(N)$ global symmetry. The background gauge fields $C_p$ couple to the $U(1)^{N-1}\subset SU(N)$ Cartan subalgebra of the flavour symmetry.

%The way we built the theory out of $N$ base pairs  and gauging a single $U(1)$, the theory comes with a $U(1)^{N-1}$ global "flavor" symmetry. In this special case, by inspection of the Lagrangian we can see that the global symmetry in theory A is in fact enhanced to a full $SU(N)$ non-Abelian flavor symmetry rotating the $N$ fermions into each other. The manifest $U(1)^{N-1}$ symmetry is the Cartan subalgbebra of this non-Abelian flavor symmetry. One may wonder how the CS level comes into play, how it affects the spectrum and what changes if we move it away from the special value of $-N$/2.

\para
 The global $U(1)$ is the topological symmetry and couples to the background field $\tilde{C}$. The basic charged object is a monopole. The coupling in the action \eqn{ffinal} takes the form $-NS_{BF}[\tilde{a},\tilde{C}]$, which tells us that a single monopole carries charge $N$ under this topological symmetry.  The properties of monopole operators have been well understood in recent years, starting from the work of  \cite{Borokhov:2002ib}; a particularly detailed account can be found in \cite{Dyer:2013fja}. The state-operator map means that one can determine the dimension of the monopole operator by considering the energy of the state on  ${\bf S}^2$ with one unit of magnetic flux. In this background, each of the $N$ electrons has a spinless fermi zero mode. This naively forms a Fock space of dimension $2^N$. By charge conjugation, these states are symmetrically spaced around zero. This means that the  Fock vacuum $|0\rangle$ has electric charge $-N/2$, while the state with  all zero modes excited has electric charge $+N/2$.

 \para
 Not all of these $2^N$ states are physical. The true ground state is dictated by Gauss' law, and this is where the Chern-Simons term plays a key role. When the Chern-Simons term has level $k$, the ground state of a single monopole must have electric charge $k$. For us, $k=-N/2$, and there is a unique monopole ground state which is simply the Fock vacuum $|0\rangle$ without any excited zero modes, which implies that the monopole is an $SU(N)$ singlet.

\para
In addition to the monopole states, Theory A also has mesons. Clearly there are $N(N-1)$ of these states, transforming in the anti-symmetric representation of $SU(N)$. Under the $U(1)^{N-1}$ Cartan subalgebra, these states have charges $(+1,-1)$ and are neutral under the topological  $U(1)$ symmetry.

%of these with non-vanishing flavor charges. Under the $U(1)^p$ Cartan elements of $SU(N)$, these charges are
%\beq
%S_i^p-S_j^p \mbox{ for any } i \neq j .
%\eeq
%The mesons are all neutral under the topological symmetry.

\para
We now turn to Theory B: the $U(1)^{N-1}$ bosonic quiver. The theory has manifest $U(1)^{N-1}$ topological symmetries. The duality tells us that these must be the Cartan sub-algebra of a full $SU(N)$ symmetry. Theory B also has a $U(1)$ flavour symmetry under which each boson has charge $+1$. This is mapped to the topological symmetry of Theory A.

\para
As explained above, when talking of flux quantisation it is prudent to work with the $U(1)^{N+1}$ form of Theory B given in \eqn{fullbaction}. Setting background fields to zero, the equations of motion for the gauge fields read
\be
 *j_i  +  \frac{f_i}{2\pi}  + \frac{\tilde{f}}{2\pi} &=& 0 \ \ \ \ {\rm and}\ \ \  \sum_i f_i =0. \label{maxeom}\ee
In this notation, the monopole of Theory A is also a monopole in Theory B, in the sense that $\int \tilde{f}=2\pi$. We can solve the equations above by setting all $f_i=0$ and turning on a charge for each current $j^i$. In the language of the $U(1)^{N-1}$ quiver theory, which we get after integrating out $\tilde{a}$, this becomes the gauge invariant state formed by the  product of all scalars, $\prod_i \phi_i$. This product state is sometimes referred to as a ``baryon" in these Abelian quivers. This is neutral under the $U(1)^{N-1}$ topological symmetries and carries flavour charge $N$. These are the same quantum numbers as the monopole in Theory A.

\para
Our next task is to look at ${f}_i$ monopoles in  Theory B.  We can construct monopole states consistent with Gauss' law constraints above  by choosing $f_j=+1$ and $f_k=-1$ for any pair $j$-$k$ of $U(1)$ fields. These have charges $(+1,-1)$ under the ultimate $U(1)^{N-1}/{\bf Z}_{N-1}$ global symmetry, matching the charges of the mesons in Theory A.  The duality predicts that this should be  enhanced to a full $SU(N)$ global symmetry group.

\subsubsection*{Predictions at Large $N$}

As we mentioned above, in general both sides of the duality are strongly coupled and we do not have control over the dimensions of operators. One important exception to this statement is the large $N$  expansion of QED. This has been much studied in recent years \cite{tappel, jsbhat,Klebanov:2011td,Giombi:2015haa}.
 To leading order, the dimension of the QED monopole operator is given by the zero-point energy in this background and evaluates to $0.265 N$ \cite{Borokhov:2002ib}. (This value is independent of the Chern-Simons level $k$ as long as $|k| \leq N/2$ so that only zero modes are needed to satisfy Gauss' law.)

\para
This operator matching gives interesting predictions  for the ``baryon" operator in the large $N$ limit of the quiver Theory B.  The large number of Abelian gauge group factors is not a standard large $N$ limit, but it would be interesting to see if there is some novel perturbative expansion possible in this theory.

\subsection{Shifting the Chern-Simons Terms}

The dualities described above inherit their Chern-Simons terms from the background, contact interaction of the seed duality \eqn{dual1}. This means, for example, that the QED theory described in Section \ref{qedsec} has Chern-Simons level fixed to be $k=-N/2$, while the  fermionic theory with multiple gauge fields has Chern-Simons levels fixed to be $-\frac{1}{2}\sum_i R_i^a R_i^b$.  These Chern-Simons levels are, however, not the most general assignments that can be made.

\para
One obvious way to generalize the dualities is to simply add an extra Chern-Simons coupling before gauging. The resulting dual theories are somewhat  cumbersome.  Nevertheless, it is instructive to see how the procedure changes the dualities.  In Section \ref{evenmoresec}, we will find a different way to alter the Chern-Simons coupling which results in more elegant dual theories.

\para
We return to our original discussion of  Abelian dualities by taking $N$ copies of \eqn{dual1}. We decompose the background fields as \eqn{adecompose} and, before gauging a $U(1)^r$ symmetry, we add the Chern-Simons term

\beq
S  = l^{ab} S_{CS}[\tilde{a}_a;\tilde{a}_b],
\nn\eeq
with the assumption that $l^{ab}$ is invertible. The result is that the Chern-Simons terms of the $U(1)^r$  fermionic theory \eqn{ffinal} are shifted to
\beq
k^{ab} = - \frac{1}{2} \kappa^{ab} + l^{ab},
\eeq
where $\kappa^{ab}$ is given by \eqn{kappaab}. In particular, we can now consider the QED theories of Section \ref{qedsec} with Chern-Simons coupling $k=-N/2+ l$.

\para
In the bosonic theory, the $U(1)^{N+r}$ gauge theory described by \eqn{fullbaction} after shifting the Chern-Simons terms becomes
\begin{eqnarray}
S_B &=& \sum_i S_0[\phi_i] +S_{\rm int}[j^a a_a]+S_{\rm int}[j^p a_p]
 + \kappa^{ab} S_{CS}[a_a; a_b]  + \kappa^{pq} S_{CS}[a_p; a_q] \nn \\&&\ \ \ \ \ \ \ \ \ \ \ \ \
 +\,  \kappa^{ab} S_{BF}[a_a -\tilde{C}_a;\tilde{a}_b] + \kappa^{pq} S_{BF}[ a_p; C_q] + l^{ab}S_{CS}[\tilde{a}_a,\tilde{a}_b].
\label{modfullbaction} \ee
Correspondingly, the equations of motion for $\tilde{a}_a$ and $a_a$ now read
\begin{eqnarray}
*j^a + \frac{\kappa^{ab}}{2 \pi} (f_b +  \tilde{f}_b) = 0   \ \ \ {\rm and}\ \ \
 \kappa^{ab}(f_b  - dC_b) + l^{ab} \tilde{f}_b = 0.
\label{modnewtwo}\end{eqnarray}
These are to be contrasted with \eqn{newtwo}: the extra $l^{ab}\tilde{f}_b$ term means that the solution to the second equation is no longer simply  $f_b = dC_b$. Instead, if we try to eliminate both $a_a$ and $\tilde{a}_b$ using their equations of motion,  we find a direct, non-local current-current interactions. Thus, there is no useful dual  in terms of the $N$ matter fields coupled to $N-r$ gauge fields.

\para
 We can, however, still integrate out $\tilde{a}_a$ to obtain a dual description in terms of $N$ gauge fields. We do this  by solving the second equation in \eqn{modnewtwo} by
\beq
\label{solvebiga} \tilde{a}_a = l_{ab} \kappa^{bc} (C_c-a_c),
\eeq
where $l_{ab}$ is the inverse of $l^{ab}$. Plugging back into \eqref{modfullbaction}, we get a theory of $N$ fermions coupled to $N$ dynamical gauge fields $a_a$ with action
\begin{eqnarray}
S_B &=& \sum_i S_0[\phi_i] +S_{\rm int}[j^a a_a]+S_{\rm int}[j^p a_p]
 + \kappa^{ab} S_{CS}[a_a; a_b]  + \kappa^{pq} S_{CS}[a_p; a_q] \nn \\&&\ \ \ \ \ \ \ \ \ \ \ \ \
 + K^{ab} S_{CS}[C_a-a_a;C_b-a_b]+ \kappa^{pq} S_{BF}[ a_p; C_q]
 - K^{ab} S_{BF}[C_a; C_b-a_b],\nn \ee
%
%\begin{eqnarray}
%\label{modfullbaction}
%S_B &=& \sum_i S_{kin,s}[\phi_i] +S_{int}[j^a a_a]+S_{int}[j^p a_p] \\&& \nonumber
% + \kappa^{ab} S_{CS}[a_a; a_b]  + \kappa^{pq} S_{CS}[a_p; a_q] + K^{ab} S_{CS}[C_a-a_a;C_b-a_b] \\&&
% +  \kappa^{ab} S_{BF}[a_a ;A_b] + \kappa^{pq} S_{BF}[ a_p; A_q]
% - K^{ab} S_{BF}[C_a; C_b-a_b]
%\end{eqnarray}
%
where $K^{ab} = \kappa^{ac} \kappa^{bd} l_{bd}$.

\subsection*{More Operator matching}

With the dual theories modified, we should revisit the matching of quantum numbers of operators. Here we focus on the QED with $N$ flavours with Chern-Simons level $k = -N/2+l$. For $0\leq l \leq N$, the monopole operator can satisfy the Gauss law constraint if we turn on $l$ zero modes. (For $l$ outside this window, we are obliged to also excite non-zero modes.) This means that for $l=1$, the monopole transforms in the fundamental representation of the $SU(N)$ flavour symmetry; for $l=2$, it transforms in the anti-symmetric representation; for higher $l$, it transforms in the $l^{\rm th}$ anti-symmetric representation.

\para
How do these transformation properties manifest themselves in Theory B? In the framework of \eqn{modfullbaction} with $U(1)^{N+1}$ gauge groups, the Gauss' law constraints are now
\be *j_i  +  \frac{f_i}{2\pi}  + \frac{\tilde{f}}{2\pi} &=& 0 \ \ \ \ {\rm and}\ \ \
\sum f_i + l\tilde{f} = 0.
\nn\ee
%
%
%As we reviewed above, the change in CS terms has dramatic effects on the spectrum. In particular, for the special case of QED with $N$ flavors, changing the CS term from $-N/2$ to $-N/2+\Delta$ means that the monopole operator needs $\Delta$ zero modes excited and therefore transforms non-trivially under the flavor symmetries. In order to not run into any trouble with flux quantization, let us first work out the spectrum in the full parent theory with $N+1$ gauge fields. Compared to case with CS level $-N/2$ only the second equation in the equations of motion \eqref{maxeom} gets modified. In the absence of
%background fields it now reads
%
%\be *j_i  +  \frac{f_i}{2\pi}  + \frac{\tilde{f}}{2\pi} &=& 0 \ \ \ \ {\rm and}\ \ \
%\sum f_i + l\tilde{f} = 0
%\nn\ee
%
We learn that when $\int \tilde{f}=2\pi$ (i.e. the simplest monopole of Theory A), it is not longer consistent to set the $f_i$ fluxes to vanish.
Instead, we must turn on $l$ $f_i$ fluxes. These carry the $U(1)^{N-1}$ global charges, mimicking the excitations of zero modes that we saw in Theory A.  In contrast, the matching of the meson states (with $\tilde{f}=0$) is unaffected.

%%%%%%%%%%%%%%%%%%%%%%%%%%%%%%%%%%%%%%%%%%%%%%%
%%% DESCRIPTION OF FLAVOUR CHARGES IN U(1)^{N-1} QUIVER THEORY
%%%%%%%%%%%%%%%%%%%%%%%%%%%%%%%%%%%%%%%%%%%%%%%%
%\para
%In this case we can also understand the spectrum once we integrated out $\tilde{a}$, but this time we need to keep track of the modified flux quantization. Recall that we solved $\tilde{a}$ via \eqref{solveforainqed}, which in terms of the extended variables reads
%\beq
%\tilde{a}=\frac{a}l}(C - \sum_i a_i).
%\eeq
%Since $\tilde{a}$ is a standard quantized field, this equation of motion implies that we need to impose a flux quantization condition that states that the sum of all fluxes $f_i$ is in an integer multiple of $l$. The gauss law following from the $a_i$ reads:
%
%\beq
%*j_i (2 \pi) = \frac{\sum f_j}{l} - f_i
%\eeq
%
%We also see from the BF coupling that the monopole number is: $N \sum f_i/l$ and the flavor charges are $f_i - f_j$.
%Now we can recover the spectrum as long as we keep in mind that the $f_i$ fluxes have to add up to an integer multiple of $l$. The simplest case is if all fluxes add to zero. Like before, one can have +1 on one node and -1 on another to recover the mesons. The single monopole maps to a flux configuration in which $sum f_i = l$. Tt manifestly carries $l$ flavor charges, just like it should.

\section{Time Reversal: Even More Abelian Duals}\label{evenmoresec}

In the previous section, we derived a class of Abelian duals starting from the seed duality
\be  Z_{\rm fermion}[A] \, e^{-\frac{i}{2}S_{CS}[A]}  = \int {\cal D} a\  Z_{\rm scalar}[a]\,e^{i S_{CS}[a] + i S_{BF}[a;A]}.\label{dual7}\ee
The action of time reversal on this duality is rather special. In the absence of background fields, the left-hand-side manifests time reversal invariance while the right-hand-side does not appear to share this symmetry. If the duality holds, time reversal must appear as a quantum symmetry on the right-hand side. This immediately gives us a second version of the duality,
\be  Z_{\rm fermion}[A] \, e^{+\frac{i}{2}S_{CS}[A]}  = \int {\cal D} a\  Z_{\rm scalar}[a]\,e^{-i S_{CS}[a] - i S_{BF}[a;A]}.\label{dual8}\ee
As explained in \cite{Karch:2016sxi,Seiberg:2016gmd}, there is some power in using both \eqn{dual7} and \eqn{dual8} together. In particular, the pair can be used to derive both the original  bosonic particle-vortex duality of  \cite{peskin,dh} and the more recently discovered fermionic particle-vortex duality of \cite{Son:2015xqa,wang1,metlitski1}.

\para
In this section, we will describe a class of dual pairs which arise from combining both forms of the duality. We start by taking  $N_L$ copies of the duality \eqn{dual7}, together with $N_R$ copies of its time reversed partner \eqn{dual8} with $N=N_L+N_R$ fermions in total.

\para
We now gauge a $U(1)^r$ symmetry. As in Section \ref{moresec}, we assign gauge charges $R_i^a$ to the fermions, where $a=1,\ldots,r$ and $i=1,\ldots,N$. We again take $R_i^a$ to have maximal rank. The remaining $U(1)^{N-r}$ global symmetries have the orthogonal charges
\be
\sum_{i=1}^N R_i^a \,S_i^p =0\ \ \ \ \ \forall\ a=1,\ldots,r\ {\rm and}\ p=1,\ldots,N-r.
\label{rsnew}\ee
The background gauge fields are decomposed as
\beq
A_i = R_i^a \tilde{a}_a + S_i^p C_p.\nn\eeq
where $\tilde{a}_a$ are the newly dynamical gauge fields and $C_p$ are the remaining background gauge fields.  So far our discussion has mirrored that of Section \ref{moresec}. However, when we insert this ansatz into our Lagrangian, the differing contact terms in \eqn{dual7} and \eqn{dual8} give different Chern-Simons terms in the resulting action, which takes the form
\be S_A &=&  i\bar{\psi}^i\!\delslash \psi^i + J^a\tilde{a}_a + J^pC_p  -\frac{1}{2} \bar{\kappa}^{ab}S_{CS}[ \tilde{a}_a; \tilde{a}_b] \nn\\ &&
\ \ \  - \frac{1}{2} \bar{\kappa}^{pq} S_{CS}[C_p; C_q] -  \bar{\kappa}^{ap} S_{BF}[\tilde{a}_a; C_p]  -\kappa^{ab} S_{BF}[\tilde{C}_a; \tilde{a}_b].
\label{fffinal}\ee
As in Section \ref{moresec}, we have introduced gauge currents $J^a=R^a_iJ^i$ and global currents $J^p=S^p_iJ^i$.  The Chern-Simons terms are now given by $k^{ab}=-\bar{\kappa}^{ab}/2$ where
\beq
\bar{\kappa}^{ab} = \sum_{i=1}^{N_L} R_i^a R_i^b - \!\sum_{i=N_L+1}^N R_{i}^a R_{i}^b \, \equiv \, \eta^{ij} R_i^a R_j^b.
\nn\eeq
Here $\eta^{ij}$ is the diagonal matrix with the first $N_L$ entries $+1$ and the remaining $N_R$ entries  $-1$.
Similarly, the other couplings in \eqn{fffinal} are
\beq
\bar{\kappa}^{pq} = \eta^{ij} S_i^p S_j^q \ \ \ {\rm and}\ \ \ \bar{\kappa}^{ap} = \bar{\kappa}^{pa} = \eta^{ij} R_i^a S_j^p.
\nn\eeq
Note that this final coefficient was absent in our original derivation \eqn{ffinal} by virtue of \eqn{rs}.  Finally, in the last term in \eqn{fffinal}, we chose to couple the background fields $\tilde{C}_a$ for the $U(1)^r$ topological currents through the original combination $\kappa^{ab}\tilde{a}_b =  R^a_i R^b_i\tilde{a}_b$.
This is because $\kappa^{ab}$ is invertible while it is possible that $\bar{\kappa}^{ab}$ has vanishing eigenvalues.
The upshot is that we have the fermionic theory
\para
{\bf Theory A:}  $U(1)^r$ gauge theory with $N$ fermions of charge $R_i^a$ and Chern-Simons levels given by
$-\frac{1}{2}\bar{\kappa}^{ab} =-\frac{1}{2} \eta^{ij} R^a_i R^b_i$.

\para
Let's now turn to Theory B. This time, we find it useful to introduce an extra minus sign in our decomposition of the original dynamical gauge fields. Instead of \eqn{olddecomp}, we write
\be
a_i = \left \{ \begin{array}{ll} R_i^a a_a + S_i^p a_p & \mbox{ for } 1 \leq i \leq N_L \\
-R_i^a a_a - S_i^p a_p & \mbox{ for } N_L+1 \leq i \leq N \end{array}
\right .\label{bgaugedef2}\ee
Since the Chern-Simons terms on the right-hand-side of \eqn{dual7} and \eqn{dual8}  are quadratic in $a_i$, this extra sign does not affect them. It does, however, change the
BF coupling between $a_i$ and $A_i$. This has the effect of undoing the sign change that arose from time reversal. We find that the dual bosonic theory is a $U(1)^{N+r}$ gauge theory whose Lagrangian generalises  \eqref{fullbaction},
\begin{eqnarray}
S_B &=& \sum_i S_0[\phi_i] +S_{\rm int}[j^a a_a]+S_{\rm int}[j^p a_p]
 + \bar{\kappa}^{ab} S_{CS}[a_a; a_b]  + \bar{\kappa}^{pq} S_{CS}[a_p; a_q] \nn \\&&\ \ \ \ \ \ \ \ \ \ \ \ \
+\, \bar{\kappa}^{ap} S_{BF}[a_a; a_p] +  \kappa^{ab} S_{BF}[a_a -\tilde{C}_a;\tilde{a}_b] + \kappa^{pq} S_{BF}[ a_p; C_q],
\label{fullbactionwith} \ee
where, $a_a$, $a_p$ and $\tilde{a}_a$ are $N+r$ dynamical gauge fields, while  $C_p$ and $\tilde{C}_a$ are $N$ background fields.
%
%\begin{eqnarray} \nonumber
%\label{fullbactionwith}
%S_B &=& \sum_i S_{kin,s}[\phi_i] +S_{int}[j^a a_a]+S_{int}[j^p a_p] \\&& \nonumber
% + \bar{\kappa}^{ab} S_{CS}[a_a; a_b]
% + \bar{\kappa}^{pq} S_{CS}[a_p; a_q] + \bar{\kappa}^{ap} S_{BF}[a_a; a_p] \\&&
% +  \kappa^{ab} S_{BF}[a_a ;A_b]  + \kappa^{pq} S_{BF}[ a_p; A_q]
 %- \kappa^{ab} S_{BF}[C_a; A_b].
%\end{eqnarray}
The equations of motion for the  $a_a$ and $\tilde{a}_a$ gauge fields are
\begin{eqnarray}
*j^a + \frac{\bar{\kappa}^{ab}}{2 \pi} f_b + \frac{\bar{\kappa}^{ap}}{2 \pi} f_p+ \frac{\kappa^{ab}}{2 \pi} \tilde{f}_b =0\ \ \ {\rm and}\ \ \
 \kappa^{ab}(f_b  - d\tilde{C}_b) = 0.
\label{newtwo2}
\end{eqnarray}
While $\bar{\kappa}^{ab}$ in general is not invertible, $\kappa^{ab}$ is. The second equation therefore enforces $a_a=\tilde{C}_a$, while the first equation allows us to eliminate $\tilde{a}_a$ by
\be d\tilde{a}_a = -\kappa_{ab} \,(2 \pi  \,  *j^b +  \bar{\kappa}^{bc} d\tilde{C}_c + \bar{\kappa}^{bp} f_p  ).
\nn\ee
Plugging this  back into the action, the $S_{BF}[a_a; \tilde{a}_b]$ and $S_{BF}[C_a; \tilde{a}_b]$ terms still cancel and we are left with
\be
S_B &=& \sum_i S_0[\phi_i] +S_{\rm int}[j^a C_a]+S_{\rm int}[j^p a_p]
 + \bar{\kappa}^{pq} S_{CS}[a_p; a_q] \nn\\ &&\ \ \ \ \ \ \ \ \ \  +\,\bar{\kappa}^{ap}S_{BF}[a_p;\tilde{C}_a] + \bar{\kappa}^{ab} S_{CS}[\tilde{C}_a; \tilde{C}_b]
 + \kappa^{pq} S_{BF}[ a_p; C_q].
 \label{bbbfinal}
\ee
In words, we have the bosonic theory:
\para

{\bf Theory B:}  $U(1)^{N-r}$ gauge theory with $N$ Wilson-Fisher scalars of charge $S_i^p$ and Chern-Simons levels
$\bar{\kappa}^{pq}=\eta^{ij} S^p_i S^q_j$.

\subsection{An Example: QED Again}\label{moreqedsec}

Returning to our favourite example of QED from Section \ref{qedsec}, we can immediately see the effects of the above time reversal procedure. We again take $r=1$ with $R_i^1=1$ for $i=1,\ldots,N$ such that the fermionic theory is now
\para
{\bf Theory A:} $U(1)$ with $N_f=N$ fermions and Chern-Simons level $k=(N_R-N_L)/2$.
\para

Meanwhile, the bosonic theory \eqn{bbbfinal} takes the form
\para
{\bf Theory B:} $U(1)^{N-1}$ coupled to $N$ Wilson-Fisher scalars with charges $S^i_p$ given by \eqn{quivers} and  Chern-Simons matrix
\be \bar{\kappa}^{pq} = \left(\begin{array}{ccccccc} 2 & -1 & & & & &  \\ -1 & \ddots & & & & & \\ & & 2 & -1 & & & \\ & & -1 & 0 & + 1 & & \\ & & & +1 & - 2 & & \\ & & & & & \ddots & +1 \\ & & & & & +1 & -2 \end{array}\right),\nn\ee
where the zero sits on the $N_L^{\rm th}$ diagonal element. In particular, in the case $N_L=N$  there is no zero on the diagonal, and $\bar{\kappa}^{pq}$ reverts to the Cartan matrix \eqn{cartan}.

\subsubsection*{Operator Matching: A Puzzle}

Theory B above offers an elegant dual to QED with modified Chern-Simons levels. It would be nice to match the operator spectrum in the same way that we saw in Section \ref{matchingsec}. Disappointingly, we have been unable to do this. We do not understand the resolution to this issue and leave it as an open problem. Below, we sketch the difficulty in constructing a working dictionary between the dual theories.

\para
Before we proceed, there is a slightly technical issue to address: the presence of the $\bar{\kappa}^{ap}$ term in \eqn{fffinal} means that the $C_p$ do not couple solely to the $U(1)^{N-1}\subset SU(N)$ Cartan currents of the fermionic theory.  Instead, these background fields mix with the topological symmetry. In the present case with $r=1$, this mixed term is  $\bar{\kappa}^{1p} = 2\delta_{p,N_L}$. We can always rectify the problem by dividing out by this term and moving it into the bosonic dual. This shifting does not seem to effect the following discussion.

\para
As we have seen previously, the level $k$ of the Chern-Simons term in the fermionic theory dictates the number of fermi zero modes that must be excited in a monopole background. This, in turn, determines the monopole's representation under the $SU(N)$ flavour symmetry. For $k=\pm N/2$, the monopole is a flavour singlet. This corresponds to $N_L=0$ or $N_R=0$ and reduces to the situation described in Section \ref{qedsec}. In contrast, for $-N/2< k<N/2$, the monopole transforms under the $SU(N)$ flavour group. It sits in the $N_R^{\rm th}$ anti-symmetric representation. For example, when $N_R=1$ is the monopole sits in the  ${\bf N}$ of $SU(N)$; when $N_R=N-1$, the monopole sits in the $\bar{\bf N}$.

\para
How does this manifest itself in the spectrum of Theory B? As we?ve seen previously, when discussing issues of flux quantisation,  it is sensible to retreat to the $U(1)^{N+1}$ theory \eqn{fullbactionwith} before integrating out any gauge fields. The Gauss' law constraints in this theory read
\begin{eqnarray}
 *j_i  \pm \frac{f_i}{2\pi}  \pm \frac{\tilde{f}}{2\pi} =0\ \ \ {\rm and}\ \ \
 \sum_{i=1}^{N_L} f_i - \sum_{i=N_L+1}^{N} f_i = 0.
\nn\end{eqnarray}
where $i=1,\ldots,N_L$ terms has the upper $+$ sign, and $i=N_L+1,\ldots, N$ comes with  the lower $-$ sign. In particular, summing these equations tells us
\beq
\sum *j_i  = (N_R-N_L) \frac{\tilde{f}}{2\pi}.
\label{doesntwork}\eeq
As we saw before, in this presentation the monopole of Theory A also requires $\int \tilde{f} = 2\pi$ in Theory B. However, the monopole inherits flavour charge if the fluxes $f_i$ are turned on and nothing above requires this to be the case. It would appear that the solution $f_i=0$ with currents saturating the constraint \eqn{doesntwork} suffices, but in this case the monopole is a flavour singlet.

\para

As we mentioned above, we do not understand the operator map between the dual theories in this case.
Presumably, the map no longer takes the simplest operators on one side to the simplest operators on the other. It would be interesting to understand this better.

\subsection{Abelian Particle Vortex Dualities}
\label{particlevortex}

Until now, we have focussed on  bosonization dualities that relate fermionic and bosonic theories. However, as shown in
\cite{Karch:2016sxi,Seiberg:2016gmd}, we can also use these to derive boson/boson and fermion/fermion particle-vortex dualities. For bosons, this duality roughly says that $U(1)$ coupled to a complex scalar is equivalent to an ungauged Wilson-Fisher scalar,
\be \int  {\cal D} a \ Z_{\rm scalar}[a]\, e^{iS_{BF}[a;A]} = Z_{\rm scalar}[-A]\label{pv1}.\ee
For fermions, it says something similar but with a subtlety. Heuristically, one can write the duality as
\be
\int {\cal D}a\ Z_{\rm fermion}[a]\,e^{\frac{i}{2}S_{BF}[a;A]}= Z_{\rm fermion}[A],\label{pv2}\ee
but with the proviso that both $a$ and $A$ only admit even fluxes, so that  $\int f$ and  $\int F \in 4\pi {\bf Z}$. A more precise definition of this duality, in which the restriction to even fluxes is implemented by an auxiliary gauge field, was given in \cite{Seiberg:2016gmd}.

\para
It is straightforward to apply the techniques of this paper (which are really the techniques of \cite{Kapustin:1999ha}), taking either \eqn{pv1} or \eqn{pv2} as a starting point. In fact the absence of Chern-Simons terms makes the manipulations somewhat simpler. In both cases, the result is an equivalence between  $U(1)^r$ gauge theory with charges $R_i^a$ and a $U(1)^{N-r}$ gauge theory with charges $S_i^p$.

\para
In the case of $U(1)$ with $N$ fermions, the theory has an $SU(N)$ flavour symmetry.  If we put back in the Maxwell term, we can ask what becomes of this symmetry in the infra-red. For a long time, the accepted wisdom was that the symmetry is preserved for $N$ large enough while, for $N$ even and less than some critical value $N_\star$, the flavour symmetry is expected to break to $SU(N/2)\times SU(N/2)\times U(1)$. There has been some debate over the value of $N_\star$. It is usually thought that $N_\star\approx 8$. However, this has been questioned in at least two ways recently. First, computations using the monoticity of certain quantities under RG flow suggest that there may actually be a ``non-conformal window", with the non-Abelian symmetry restored again for $N=2$ \cite{Safdi:2012re}. Second, lattice simulations first using Wilson fermions \cite{Karthik:2015sgq} and subsequently overlap fermions \cite{Karthik:2016ppr} have recently suggested that the flavour symmetry is unbroken for all $N$. An up-to-date and lucid discussion of the current status of these theories can be found in \cite{Gukov:2016tnp}.

\para
This has consequence for the dual theories where only a $U(1)^{N-1}$ is manifest as topological symmetries. If the flavour symmetry of QED is unbroken in the infra-red, then the topological symmetries of the quiver theory should be enhanced to $SU(N)$. This  kind of non-Abelian symmetry enhancement is well known in supersymmetric mirror symmetries \cite{Intriligator:1996ex}

\para
The discussion above holds for the fermionic particle-vortex mirrors. For bosons, the quartic terms in the action \eqn{scalar} which drive the theory to the Wilson-Fisher fixed point mean that there is no $SU(N)$ global symmetry rotating different scalars. However, if the Legendre transform discussed in Section \ref{rgsec} is valid, then a similar discussion to that above for fermions holds.

\subsection*{A special example: QED with 2 flavors}

QED with $N=2$ flavors has a particularly nice property: it was conjectured in \cite{Xu:2015lxa} to be self-dual in the sense that one can exchange the flavour and topological symmetries. This was confirmed in  \cite{Karch:2016sxi} using the kinds of techniques described here. (See also \cite{Hsin:2016blu} for a subsequent discussion.) There we showed that
\be \nn Z_{[N_f=2]}[A;C] \equiv
 \int {\cal D}a\ Z_{\rm fermion}[a+C] Z_{\rm fermion}[a-C] \, e^{+iS_{BF}[a;A]} =  \bar{Z}_{[N_f=2]}[C;A],
 \nn\ee
where the notation $\bar{Z}$ denotes the time-reversed partition function. In the first partition function, $C$ couples to the current $U(1)\subset SU(2)$ and $A$ is the topological symmetry; in the second partition function their roles are reversed. If the $SU(2)$ flavour symmetry is unbroken in the infra-red, it must be that this theory exhibits an enhanced $SU(2)\times SU(2)$ symmetry. This self-duality can be seen as a special case of the general Abelian particle/vortex dualities we described in section \ref{particlevortex} for $N=2$, $r=1$, $R=(1,1)$ and $S=(1,-1)$. The extra sign in the second entry in $S$ can be undone by charge conjugation.

\para
There is one further surprise in this theory observed in  \cite{Karch:2016sxi}: the partition function coincides with that of QED coupled to two scalars. A relationship between these two theories was previously suggested in \cite{Alicea:2005fp,Senthil:2005jk,Vishwanath:2012tq}. Here scalar QED is defined as
\be
 Z_{[N_s=2]}[A;C] = \int {\cal D}a \ Z_{\rm scalar}[a+C] Z_{\rm scalar}[a-C] e^{iS_{BF}[a;A]} .\nn\ee
This theory too is self-dual. In  \cite{Karch:2016sxi} it was shown that
\be Z_{[N_f=2]}[A;C] = Z_{[N_s=2]}[A-C;(A+C)/2].\nn\ee
This also follows from the duals derived in Section \ref{moreqedsec}.

\para
QED with 2 flavors has recently been argued to remain self-dual when we take the charges to the two flavors to be $1$ and $k$ instead of having unit charge for both flavors \cite{cenke}. This self-duality in fact also nicely falls into our general scheme for the Abelian duals from section \ref{particlevortex}, where this time we take $N=2$, $r=1$, $R=(1,k)$ and $S=(-k,1)$. This generalized self-dual pair is particularly interesting as it allows a large $k$ limit in which one has some analytic control over the theory \cite{cenke}.

\section*{Acknowledgements}

We are supported by the US Department of Energy under grant number DE-SC0011637, by STFC, and by the European Research Council under the European Union's Seventh Framework Programme (FP7/2007-2013), ERC grant agreement STG 279943, ``Strongly Coupled Systems". DT is grateful to the Stanford Institute for Theoretical Physics and to the University of Washington for their kind hospitality while this work was undertaken. Our thanks to Ofer Aharony, Ethan Dyer,  Guy Gur-Ari,  Shamit Kachru, Charles Melby-Thompson, Djordje Radicevic and Carl Turner for useful discussions.

\bibliographystyle{JHEP}
\bibliography{mirror}

\end{document}